\documentclass[aps,prl,groupedaddress,10pt,twocolumn,reprint,amsmath,amssymb,floatfix]{revtex4-1}

\usepackage{graphicx}
\usepackage{dcolumn}
\usepackage{bm}
\usepackage[usenames, dvipsnames]{color}
\usepackage{ulem,amssymb,amsmath,color,amsbsy,alltt,natbib,graphicx,epsf}

\usepackage[T1]{fontenc}
\usepackage{lmodern,enumitem}
\usepackage{xcolor}
\usepackage[export]{adjustbox}
\usepackage{empheq}

\begin{document}

\title{Elucidating the oscillation instability of sessile drops triggered by surface acoustic waves}

\author{Nicolas Chastrette$^{1, 2}$}
\author{Micha\"{e}l Baudoin$^{3,4}$}
\author{Philippe Brunet$^{2}$}
\author{Laurent Royon$^{5}$}
\author{R\'egis Wunenburger$^{1}$}
\email{regis.wunenburger@sorbonne-universite.fr}
\affiliation{$^1$Sorbonne Universit\'e, CNRS, Institut Jean Le Rond d'Alembert, F-75005 Paris, France\\
$^2$Universit\'e de Paris, MSC, UMR 7057, CNRS, F-75013 Paris, France\\
$^3$Univ. Lille, CNRS, Centrale Lille, ISEN, Univ. Polytechnique Hauts de France, UMR 8520, IEMN, F-59000 Lille, France\\
$^4$Institut Universitaire de France, 1 rue Descartes, 75231 Paris Cedex 05, France\\
$^5$Universit\'e de Paris, LIED, UMR 8236, CNRS, F-75013 Paris, France}
\begin{abstract}
The oscillation instability of sessile drops is ubiquitous in surface acoustic wave (SAW)-powered digital microfluidics. Yet, the physics underlying these phenomena has not been elucidated owing to the interplay between hydrodynamics, acoustics and capillarity. We decipher the instability by combining high-speed imaging with pressure measurements. We rationalize the observed behaviour with a model inspired from optomechanics, which couples an intracavity acoustic mode excited by the SAW to a surface deformation eigenmode through amplitude modulation and delayed radiation pressure feedback.
\end{abstract}

\date{\today}

\maketitle
Manipulating microparticles
or fluid samples is a key issue in microfluidics 
for e.g. high throughput screening, bottom-up processing, selectivity or detection sensitivity enhancement or risk reduction~\cite{Tabelling}. 
Actuation by surface acoustic waves (SAW) has been early identified as a versatile and efficient tool in both microchannel and drop microfluidics~\cite{friend2011,yeo2014,connacher2018,ieee_riaud_2016}. The interaction of a SAW with a sessile drop results in various phenomena depending on the liquid viscosity, drop size, SAW frequency, phase and intensity distribution: drop trapping~\cite{alvarez_2008}, oscillations and transport~\cite{wixforth2003,renaudin2006,brunet2010,baudoin2012,bussonniere2016}, liquid atomisation~\cite{ieee_shiokawa_1989,jjap_shiokawa_1990,jjap_chono_2004,pof_qi_2008,epl_tan_2009}, particle transport or segregation~\cite{loc_tan_2007,rezk2014}, mixing~\cite{prl_frommelt_2008} and heating~\cite{saa_kondoh_2009,pnas_reboud_2012,afm_shilton_2015}. The 
physical mechanisms at play combine hydrodynamics,  
acoustics, capillarity and wetting. In this letter, we address the $\sim \!\! 10^2$~Hz surface oscillation instability of sessile drops insonicated by $\sim \!\! 10^6$~Hz SAWs, a counter-intuitive phenomenon that has been observed to precede surface wave turbulence~\cite{blamey2013} as well as two key-applications of SAWs in microfluidics, namely atomisation~\cite{qi2009,friend2011} and drop transport~\cite{brunet2010,baudoin2012}, and whose elucidation as been quoted as one of the major fundamental challenges in Yeo and Friend's reviews on acoustofluidics~\cite{friend2011,yeo2014}. 

When a SAW irradiates a sessile drop, it is partially converted to a bulk longitudinal wave, which remains confined within the drop acting as a cavity. Here we operate at moderate acoustic excitation frequency $f_{\text{ae}} \simeq 1$~MHz such that $\lambda/R \simeq 1$ 
($c \simeq 1.5~\text{km.s}^{-1}$ the sound velocity in water, $\lambda=c/f_{\text{ae}}$ the acoustic wavelength and $R \simeq 1.6$~mm the drop radius). 
This enables to excite a single acoustic mode in the cavity, while at higher frequencies, high modal density leads to mode overlapping and in turn to chaotic behaviour~\cite{Riau17}. 
Furthermore, 
a drop is deformable and behaves as a mechanical oscillator whose stiffness is associated to surface tension and mass to liquid inertia~\cite{Rayleigh,Lamb}. By combining high-speed imaging with acoustic pressure measurements and using an instability model inspired from optomechanics, we demonstrate that drop oscillations result from the mutual interaction between the mechanical oscillator and the confined wave field, through amplitude modulation and delayed radiation pressure feedback, a scheme reminiscent of the parametric instability of gravitational wave interferometric detectors \cite{Abramovici1992}, force sensing microlevers \cite{Jourdan2008,Favero2009} and opto-mechanical oscillators designed for quantum intrication \cite{Metzger2008}. 
\begin{figure}[h]
\centering
\centering\includegraphics[width=85mm,angle=0]{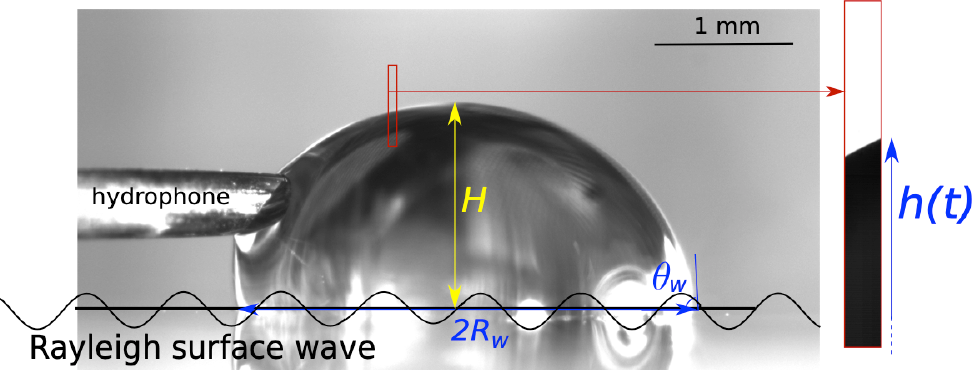}
\caption{A $10.0~\mu \text{L}$ water drop deposited on a glass slab is excited by a harmonic progressive Rayleigh wave. A needle hydrophone, whose tip is brought into contact with the liquid surface,  
measures the intracavity pressure $p(t)$. Full pictures of the slowly evaporating drop are taken at $30$~fps. 
Oscillations of the drop free surface close to its apex (red frame on the main picture)  are recorded at $1000$~fps using a high-speed camera in a narrow field (shown in inset).
\label{figure1}}
\end{figure}

A sessile water drop of volume $V=10.0~\mu \text{L}$ partially wets a $10$~mm-thick glass slab with a contact angle $\theta{_\mathrm{w}}$ close to $90^{\circ}$ and a contact line radius $R_w=1.65$~mm (see Fig.~\ref{figure1} and~\cite{SM}). The drop is irradiated by a plane, progressive Rayleigh wave~\cite{SM}. 
The tip of a needle hydrophone with $200~\mu \text{m}$-diameter active element is put in contact with the liquid, allowing for a measurement of the intracavity acoustic pressure $p(t)$. Full pictures of the drop are acquired using a camera at $30$~fps, while the oscillations of its free surface in a narrow field close to its apex (Fig.~\ref{figure1}) are recorded at $2.2~\mu\text{m/pixel}$ resolution using a high-speed camera at $1000$~fps. The experiments consist in exciting a drop with a harmonic SAW at carrier frequency $f_{\mathrm{ae}}$ close to one of the resonance frequencies $f_{\mathrm{ar}}$ of the acoustic cavity. Amplitude ramps are applied in order to detect the onset of the oscillation instability. 
\begin{figure}[h]    
    \centering
    \includegraphics[width=0.45\textwidth]{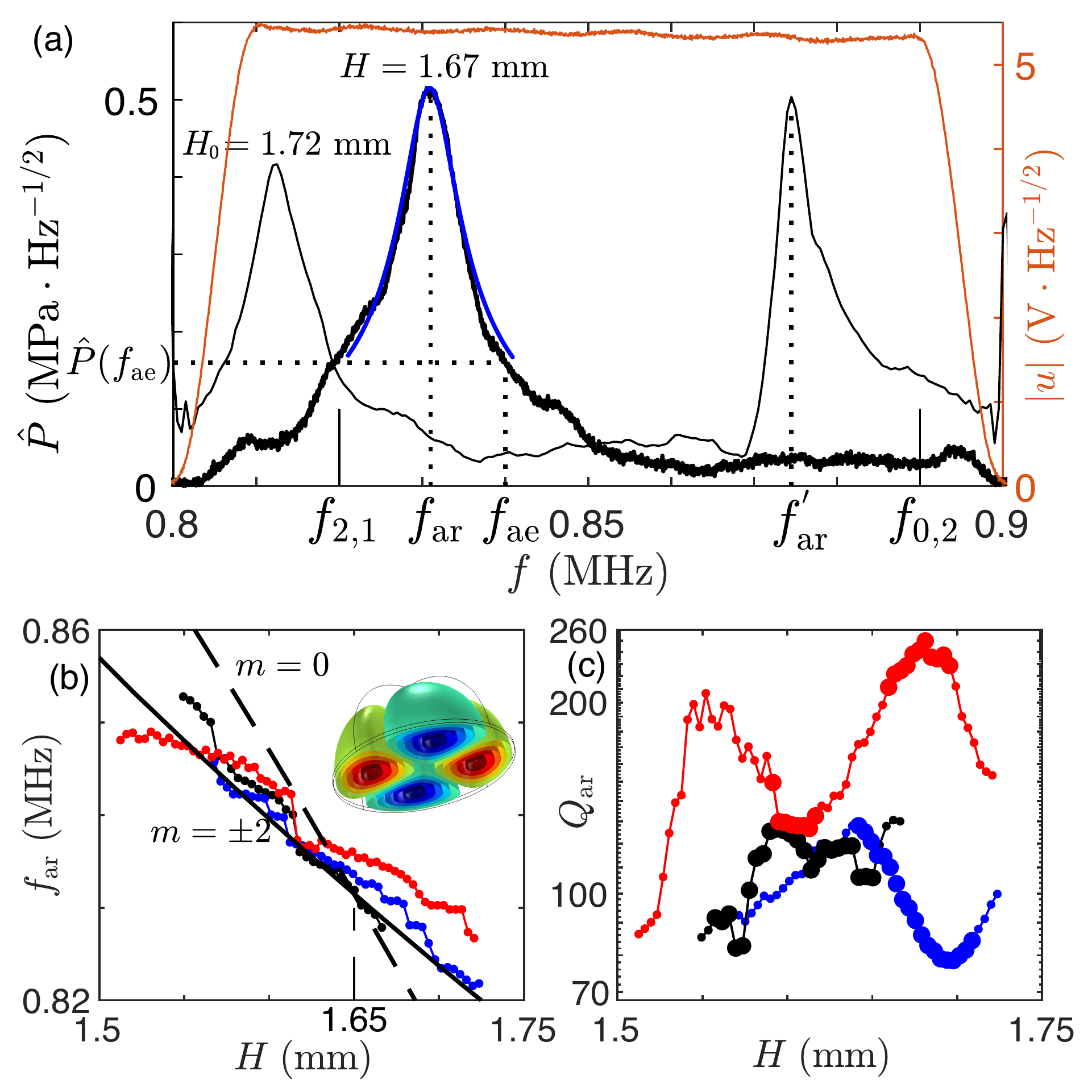}
    \caption{(a) Red curve: Narrow-band spectrum $|u(f)|$ of the frequency-swept voltage periodically applied to the transducer to scan the drop acoustic resonances during its evaporation. Spectrum $\hat{P}(f) = |p(f)/u(f)|$ of the drop pressure response immediately after its deposition (thin black curve) and after $3$~min of evaporation (bold black curve). Blue curve: best fit by Eq.~\eqref{eq:P}. $f_{2,1}$ and $f_{0,2}$ are the numerically predicted acoustic eigenfrequencies of a hemispherical drop with radius $R=1.68$~mm. (b) Acoustic resonance frequency $f_{\rm ar}$ vs $H$ during the evaporation of three drops (connected red, black, blue symbols). Solid and dashed curves: theoretical predictions, see text. Inset: pressure distribution of the forced $m=\pm 2$ eigenmodes. (c) Corresponding acoustic quality factor $Q_{\rm ar}$. Bold symbols: measurements during which the instability is observed. \label{figure2}}
\end{figure}

The linear acoustic response of the cavity formed by the drop is determined by supplying the transducer with a voltage signal $u(t)$ whose frequency linearly sweeps over the range $0.8-0.9$~MHz, with low amplitude $U=5$~V and $20$~ms duration (its spectral amplitude $|u(f)|$ is displayed in Fig.~\ref{figure2}(a)). The spectral amplitude $\hat{P}(f) = |p(f)/u(f)|$ of the pressure response of a freshly deposited drop (height $ H_0  \! = \! 1.72$~mm) is shown in Fig.~\ref{figure2}(a), where $p(f)$ is the Fourier transform of $p(t)$. $\hat{P}$ exhibits two peaks in the range $0.8-0.9$~MHz respectively at $f_{\mathrm{ar}} \simeq 0.81$~MHz and $f^{'}_{\mathrm{ar}} \simeq 0.87$~MHz. Considering the drop as hemispherical and assuming (i) no displacement of the rigid glass surface and (ii) pressure node at the drop free surface, the  acoustic eigenmodes can be expressed using spherical waves \cite{arfm_baudoin_2020} and their eigenfrequencies $f_{\ell,q}$ satisfy $j_{\ell} \left( 2 \pi Rf_{\ell,q}/c \right)=0$ where $R=(3V/(2 \pi))^{1/3}=1.68$~mm, $j_{\ell}$ is the spherical Bessel function of the first kind, $\ell$ a positive integer and $q$ the number of the root of $j_{\ell}$ in ascending order~\cite{SM}. Two eigenfrequencies $f_{2,1} = 0.82$~MHz (corresponding to $3$ degenerate modes symmetrical to the equatorial plane, labelled by $m=0,\pm 2$~\cite{SM}) and $f_{0,2} = 0.89$~MHz fall in the range $0.8-0.9$~MHz and are found to be close to the measured peak frequencies $f_{\mathrm{ar}}$ and $f^{'}_{\mathrm{ar}}$. This demonstrates that these peaks do correspond to the resonant forcing of cavity eigenmodes.

Due to evaporation, the height $H$ of the drop apex decreases in time while its contact line remains pinned during the first $5$~min, resulting in a continuous increase of $f_{\mathrm{ar}}$, see Fig.~\ref{figure2}(a, b). A numerical calculation (using {\it Comsol} software) of the acoustic eigenmodes of a sessile drop with fixed $R_w$, initial volume $V$ and initial height $H_0$ evidences that, when the drop is not hemispherical ($H \! \neq \! R_w$), the eigenfrequencies of the $(\ell = 2, q=1)$ modes split~\cite{SM}. The comparison between the variations of $f_{\mathrm{ar}}$ with $H$ measured during several experiments and the computed variations of the eigenfrequencies of the $m = \pm 2$ and $m=0$ modes versus $H$, shown in Fig.~\ref{figure2}(b), allows us to unambiguously identify $f_{\mathrm{ar}}$ as the resonance frequency of the $(\ell = 2, q=1, m = \pm 2)$ modes, whose pressure field is shown in inset of Fig.~\ref{figure2}(b).

As shown in Fig.~\ref{figure2}(a), during evaporation the pressure response is accurately described by a one-dimensional (1D) resonator model: 
\begin{equation}\label{eq:P}
    \hat{P}(f)=\hat{P}_{\text{i}} \left| 1+ \left[1 - \frac{\pi}{2Q_{\text{ar}}} \exp \left(i \pi \frac{f}{f_{\text{ar}}}\right) \right]  \right|^{-1}
\end{equation}
where $\hat{P}_{\text{i}} = (3.5 \pm 1.5)~\text{kPa} \cdot \text{V}^{-1}$ is the magnitude related to the excitation of the acoustic wave refracted through the drop by mode conversion and $Q_{\text{ar}}$ is the quality factor. Fig.~\ref{figure2}(c) displays the variations vs $H$ of $Q_{\text{ar}}$ measured over several experiments. Its fluctuations lie in the range $80-250$ and may be ascribed to the slight variations of the triple line shapes along the substrate and needle tip from one experiment to another.

\begin{figure}[ht!]
    \centering
    \includegraphics[width=0.45\textwidth]{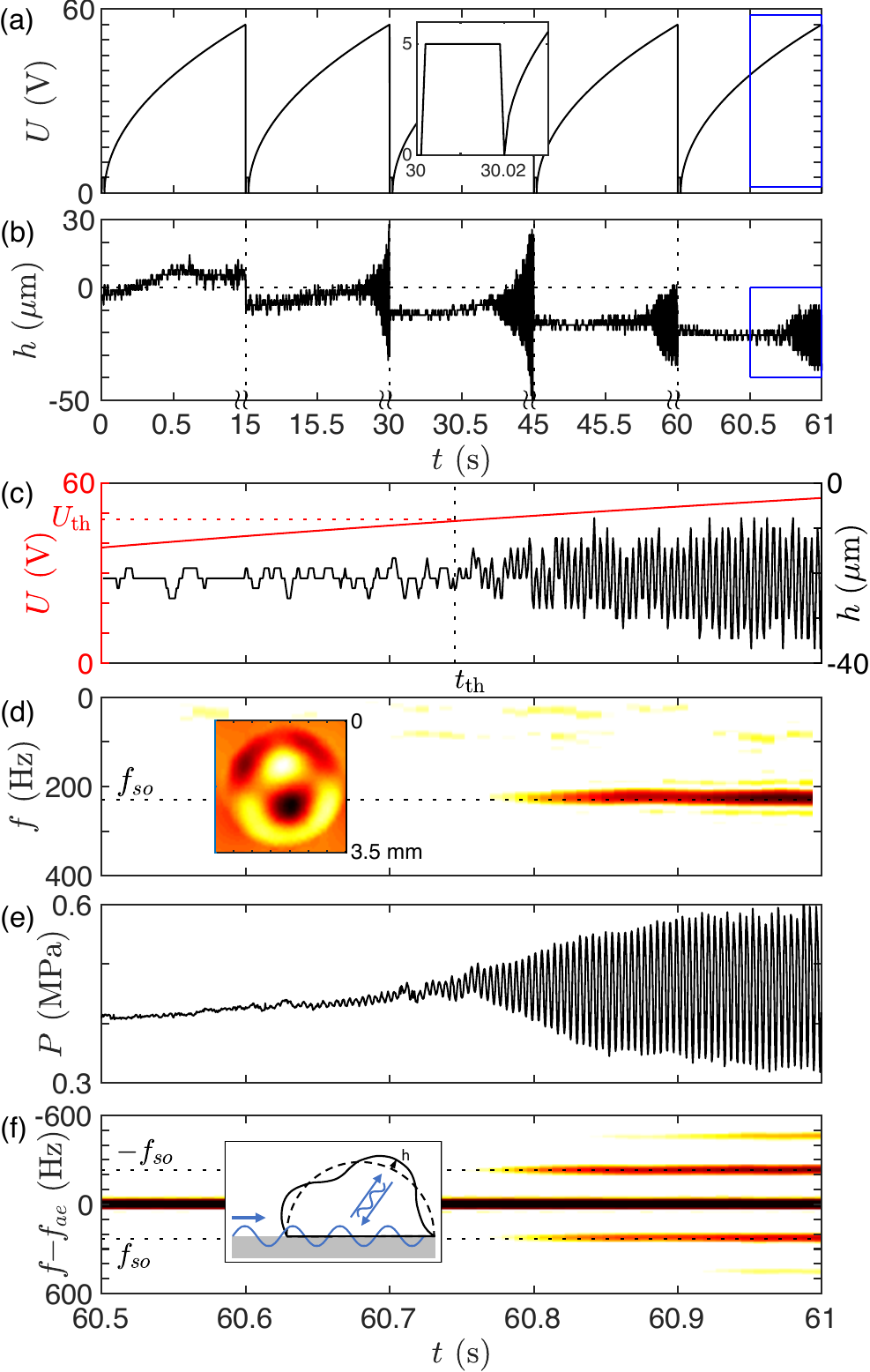}
    \caption{(a) Amplitude $U(t)$ of the sine voltage signal prescribed to the transducer, which is 15~s-periodic, 1~s-long (note the breaks on time scale). Inset: zoom on the frequency-swept, small-amplitude signal used to assess the cavity acoustic response. (b) Relative height $h(t)$ (arbitrary origin) of the drop surface imaged using the high-speed camera, undergoing a slow decrease due to evaporation and possibly oscillations during excitation. (c) Enlargements of the parts of the signals framed in blue in (a) and (b), evidencing the surface oscillation instability triggered for $U  \geq U_{\rm th}$. (d) Spectrogram of $h(t)$ evidencing a monochromatic oscillation at frequency $f_{\rm so} \simeq 230$~Hz. Inset: corresponding drop deformation pattern measured from above~\cite{SM}. (e) Amplitude $P(t)$ of the intracavity pressure $p(t)$, displaying modulation starting from $t_{\rm th}$. (f) Spectrogram of $p$ showing the appearance of satellite peaks shifted from the carrier frequency $f_{\rm ae}$ by $\pm f_{ \rm so}$ starting from $t_{\rm th}$. Inset: sketch of the feedback of the drop deformation on the intracavity field.
    \label{figure4}}
\end{figure}

We now address the oscillation instability. Due to evaporation, $f_{\mathrm{ar}}$ increases at a typical rate of $3.5~\text{kHz.min}^{-1}$. Instead of continuously adjusting the frequency $f_{\text{ae}}$ of the acoustic excitation for maintaining the same interference conditions in the drop, we set $f_{\text{ae}}$ slightly above the initial value of $f_{\mathrm{ar}}$ (see Fig.~\ref{figure2}(a)) i.e. $f_{\text{ae}}=0.84$~MHz. Hence, $f_{\mathrm{ar}}$ crosses $f_{\mathrm{ae}}$ after $3\! - \! 4$~min and the drop acoustic resonance is scanned in $5$~min, which leads us to analyze how the drop unstable behavior changes with time. Accordingly, every $15$~s, (i) the linear acoustic response of the cavity is measured as described above, (ii) then the drop is insonicated using a harmonic SAW with fixed frequency $f_{\mathrm{ae}}$, amplitude $U$ increasing as $\sqrt{t}$ from $0$~V to $55$~V in $1$~s, as sketched in Fig.~\ref{figure4}(a), so that the radiation pressure (RP) exerted by the intracavity acoustic field on the drop surface, which is proportional to $U^2$, increases linearly in time, (iii) while $p(t)$ and the drop surface height $h(t)$, defined in Fig.~\ref{figure1} and shown in Fig.~\ref{figure4}(b), are recorded, (iv) the remaining time being dedicated to data transfer.

\begin{figure}[h]
    \includegraphics[scale=1]{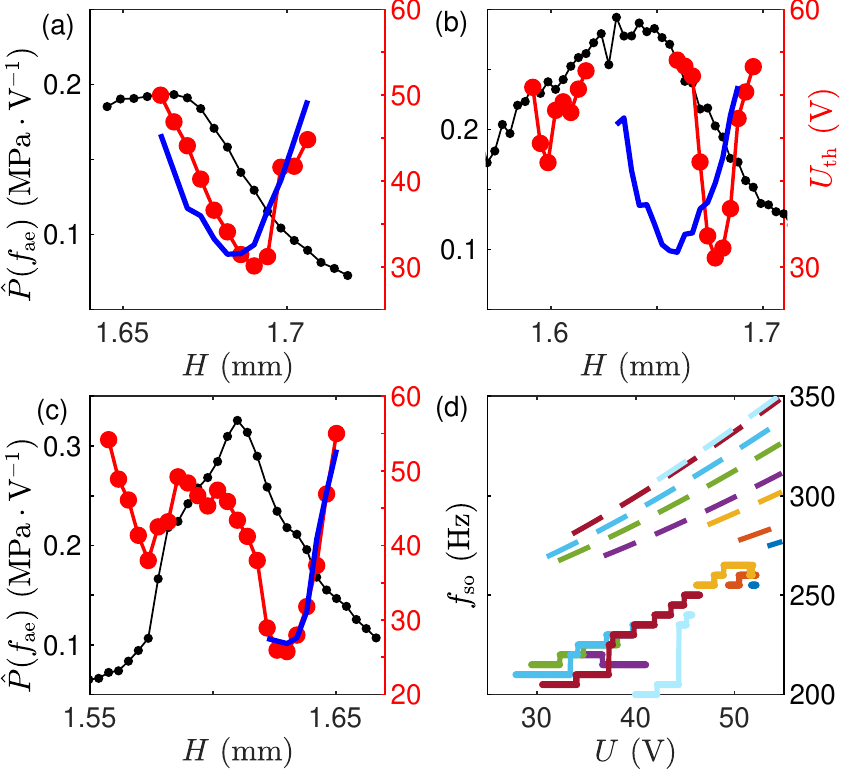}
    \caption{(a-c) Left scale, joined black symbols: $H$-dependent pressure response $\hat{P}$ at excitation frequency $f_{\mathrm{ae}}$. Right scale, joined red symbols: instability threshold voltage $U_{\rm th}$ vs $H$. Blue curve: best fit of $U_{\rm th}(h)$ by Eqs. (\ref{eq:Kasurm}, \ref{eq:seuil}). (d) Solid curves: measured surface oscillation frequency $f_{\text{so}}$ vs $U$ ($\geq U_{\text{th}}$) for several values of $H$. Dashed curves: corresponding predictions from Eq.~\eqref{eq:omega}.
    \label{figure3}}
\end{figure}

As it can be understood from Fig.~\ref{figure2}(a), during evaporation, starting from non-resonant conditions (``blue detuning'' in optical physics), the drifting resonance frequency $f_{\mathrm{ar}}$ crosses $f_{\mathrm{ae}}$ for a given height $H^{\text{res}} \simeq 1.63$~mm, realizing a resonant forcing. When $f_{\mathrm{ar}}$ overruns $f_{\mathrm{ae}}$, the cavity is driven away from resonance (``red detuning''). Accordingly, as shown in Figs.~\ref{figure3}(a-c), when $H$ decreases, the magnitude of the intracavity pressure response to excitation $\hat{P}(f_{\text{ae}})$ first increases, then reaches a maximum when $f_{\mathrm{ar}}=f_{\mathrm{ae}}$, and finally decreases. The effect of the evaporation on the drop eigenmode excitation and the way to take advantage of it are reminiscent of a previous study on concentration patterns of colloids in SAW-excited drops~\cite{Li_PRL2008}.

Meanwhile, as shown in Fig.~\ref{figure4}(b, c), $h$ slowly decreases and undergoes oscillations during excitation when $U$ exceeds a threshold $U_{\mathrm{th}}$. The variations of $U_{\mathrm{th}}$ vs $H$, measured during several experiments, some of which displayed in Fig.~\ref{figure3}(a-c), reveal that two instability tongues reproducibly show up on the right-hand side ($f_{\mathrm{ae}} > f_{\mathrm{ar}}$) and left-hand side ($f_{\mathrm{ae}} < f_{\mathrm{ar}}$) of the peak on the curve of variation of $\hat{P}(f_{\text{ae}})$ vs $H$,  
demonstrating that 
the instability occurs close to (and not at) an eigenfrequency of the acoustic cavity. Besides, 
Fig.~\ref{figure2}(c) shows that the occurrence of the instability is not correlated to the variations of $Q_{\text{ar}}$.

The spectrogram of $h(t)$ shown in Fig.~\ref{figure4}(d) evidences that, at the instability onset, $h$ shows harmonic oscillations of frequency $f_{\mathrm{so}} \simeq 230$~Hz. An independent recording of the deformations of the image of a grid visualized through the drop~\cite{SM} allowed us to measure the free-surface deformations and to identify the corresponding forced surface eigenmode, whose experimental pattern is shown in inset of Fig.~\ref{figure4}(d)~\cite{SM,bostwick2014}. We conclude that the instability 
involves a single surface eigenmode. 

To gain insight into the instability mechanism, we now consider the time evolution of the amplitude $P$ of $p(t)$, shown in Fig.~\ref{figure4}(e): $P$ smoothly increases with $U$ up to the instability threshold. Once the surface oscillates, $P$ exhibits oscillations at $f_{\mathrm{so}}$, whose amplitude increases with $U$. This modulation can be explained as follows: the drop behaves as a 1D resonator with an effective length modulated by the free surface oscillations $h(t)  = A \cos{(\omega_{\mathrm{so}}t)}$ ($\omega = 2 \pi f$ throughout the text), see inset in Fig.~\ref{figure4}(f). Hence, $f_{\text{ar}}$ is modulated at frequency $f_{\text{so}}$ with amplitude $A \frac{d f_{\text{ar}}}{dh}$,  
where $\frac{d f_{\text{ar}}}{dh}$, which quantifies the sensitivity of $f_{\text{ar}}$ to drop deformations, is negative since a 1D resonator eigenfrequency decreases with its length. Since the delay of adaptation of $p(t)$ to changes in interference conditions $\tau_{\text{ar}} = Q_{\text{ar}} / \omega_{\text{ar}} \simeq 20~\mu \text{s}$ is much shorter than their period $f_{\text{so}}^{-1} \simeq 4$~ms, $P(t)$ quasistatically follows these changes and in turn adopts a similar modulation at frequency $f_{\text{so}}$, with amplitude $\Delta P = U \frac{\partial \hat{P}}{\partial f_{\text{ar}}}(f_{\text{ae}}) A \frac{d f_{\text{ar}}}{dh}$. This is experimentally confirmed by the spectrogram of $p$ shown in Fig.~\ref{figure4}(f), which displays a pair of satellite peaks shifted from $f_{\text{ae}}$ by $\pm f_{\mathrm{so}}$ \cite{SM}.

Finally, we address the feedback of the amplitude-modulated intracavity field on the surface oscillations. The aforementioned RP, defined as the time-average over $f_{\text{ae}}^{-1}$ of the pressure $p(t)$ exerted on the drop free surface, results in a normal stress oriented outwards and scaling as $\Pi = P^2/(\rho c^2)$, where $\rho$ is the water density~\cite{Herrey1955,Borgnis1953a}. The first ingredient of the proposed feedback is that the intracavity pressure modulation amplitude $\Delta P$ induced by the surface oscillations results in RP oscillations of amplitude $\Delta \Pi = 2P \, \Delta P / (\rho c^2) $ at instability onset, which consequently have the same frequency $f_{\text{so}}$ as the surface oscillations. These RP oscillations may damp or amplify the surface oscillations depending on their phase difference. Close to resonance, the amplitude $h$ of the surface eigenmode forced by RP oscillations follows the dynamics of a forced mass-spring system: $\ddot{h}+\frac{\omega_{so}}{Q_{\text{so}}} \dot{h} + \omega_{\text{so}}^2 h = \frac{F(h)}{m}$, where $\dot{h}$ is the time derivative of $h$ and $Q_{\text{so}}= (40 \pm 15)$ is the independently measured quality factor of the resonance of the surface eigenmode~\cite{SM}. Evaluating the effective wavenumber of the surface eigenmode as $k_{\text{so}}=\frac{4}{R}$ \cite{SM}, $F \propto k_{\text{so}}^{-2} \Delta \Pi$ is the modulated acoustic radiation force exerted on a portion of drop surface of characteristic size equal to one deformation wavelength $\lambda_{\text{so}}=2 \pi/k_{\text{so}}$ and $m \propto \rho k_{\text{so}}^{-3}$ the corresponding mass of moving water.

At the instability onset, we can linearize the variations of $F$ around equilibrium: $\frac{F(h)}{m} = \frac{K_{\text{a}}}{m}h$ where $K_{\text{a}}=F^{'}(0)$ accounts for a modification of the surface stiffness induced by RP (``acoustic spring'' effect~\cite{issenmann2013}). Thus, $h$ satisfies: $\ddot{h}+\frac{\omega_{\text{so}}}{Q_{\text{so}}} \dot{h} + \left( \omega_{\text{so}}^2 - \frac{K_{\text{a}}}{m} \right) h = 0$ where
\begin{equation}\label{eq:Kasurm}
    \frac{K_{\text{a}}}{m} = \frac{ k_{\text{so}}}{\rho^2 c^2} U^2 \hat{P} \frac{\partial \hat{P}}{\partial f_{\text{ar}}}(f_{\text{ae}}) \frac{df_{\text{ar}}}{dh}
\end{equation}
within an unknown factor. First, we focus on the right-hand side of the peak of the $\hat{P}(f_{\text{ae}})$ vs $H$ curve shown in Fig.~\ref{figure3}(a-c) ($f_{\mathrm{ae}} > f_{\mathrm{ar}}$), along which $\frac{\partial \hat{P}}{\partial f_{\text{ar}}}(f_{\text{ae}}) >0$ since a decrease of $f_{\text{ar}}$ due to a crest ($h>0$) detunes the cavity, see Fig.~\ref{figure2}(a). Hence $K_{\text{a}}<0$ and the eigenmode dynamics is that of a stiffer, yet damped free harmonic oscillator
exhibiting no instability. 
A missed point is the finite delay $\tau_{\text{ar}}$ of adaptation of $P$, and hence of $\Pi$, to the changes of the intracavity interference conditions caused by the surface oscillations. In the harmonic regime, this results in a phase lag of $F$ with respect to $h$ : $K_{\text{a}}$ is to be changed to $K_{\text{a}} \exp(-i \phi)$, where  $\phi = \omega \tau_{\text{ar}} \simeq \omega_{\text{so}} \tau_{\text{ar}} \simeq 0.03 \ll 1$. Thus, in the harmonic regime, $h$ satisfies:
\begin{equation}\label{eq:osc}
\ddot{h} + \left( \frac{\omega_{\text{so}}}{Q_{\text{so}}} + \frac{K_{\text{a}}}{m \omega} \sin \phi \right) \dot{h} + \left( \omega_{\text{so}}^2 - \frac{K_{\text{a}}}{m} \cos \phi \right) h = 0.
\end{equation}
Since $\sin \phi > 0$ and $K_a < 0$, the damping term in Eq.~(\ref{eq:osc}) is reduced by the delayed feedback. The condition for instability is a negative damping occurring for $U \geq U_{\text{th}}$ such that:
\begin{equation}\label{eq:seuil}
\frac{|K_{\text{a}}(U_{\text{th}})|}{m} = \frac{\omega_{\text{ar}}\omega_{\text{so}}}{Q_{\text{ar}}Q_{\text{so}}}
\end{equation}
given $\omega \simeq \omega_{\text{so}}$ and $\sin \phi \simeq \phi$. Furthermore, as shown by Eq.~\eqref{eq:osc}, the acoustic spring effect increases the oscillation frequency $\omega^{'}_{\text{so}}$:
\begin{equation}\label{eq:omega}
\omega^{'}_{\text{so}} = 
\sqrt{\omega_{\text{so}}^2 - \frac{K_{\text{a}}}{m}} 
\end{equation}
since $\cos \phi \simeq 1$.

The instantaneous oscillation frequency $f^{'}_{\text{so}}$ is extracted from the spectrograms of $h(t)$ measured for several values of $H$. Fig.~\ref{figure3}(d) evidences the increase of  $f^{'}_{\text{so}}$ with $U$, which is in agreement with the stiffening predicted by~\eqref{eq:omega}. The negative offset of $f^{'}_{\text{so}}$ with respect to $f_{\text{so}}=230$~Hz may be ascribed to the RP-induced static deformation of the drop, see~\cite{SM}. To test the quantitative validity of this model for $f_{\mathrm{ae}} > f_{\mathrm{ar}}$, we compare Eqs.~\eqref{eq:seuil} and \eqref{eq:omega} to the experimental data using Eq.~\eqref{eq:P} fitted to the resonance curve at each value of $H$, Eq.~\eqref{eq:Kasurm} and $Q_{\text{ar}} = 100$, $\alpha$ (defined as $\frac{df_{\text{ar}}}{dh} = - \alpha \frac{f_{\text{ar}}}{H}$) being the only fitting parameter. As shown in Figs.~\ref{figure3}(a-c), the measured and predicted instability tongues $U_{\text{th}}(H)$ are in quantitative agreement. The same goes for the measured and predicted rates of variation of $f^{'}_{\text{so}}(U)$, as shown in Fig.~\ref{figure3}(d). Moreover, the values of $\alpha$ corresponding to the best fits of $U(H)$ and of the rates of variation of $f_{\rm so}(U)$ are scattered in a narrow range (between $2 \cdot 10^{-4}$ and $10^{-3}$). Thus, the model reproduces quantitatively and self-consistently two main observables of the instability, namely its threshold and the RP-induced stiffening.

Finally, we consider the left-hand side of the peak of the $\hat{P}(f_{\text{ae}})$ vs $H$ curve ($f_{\mathrm{ae}} < f_{\mathrm{ar}}$), along which $\frac{\partial \hat{P}}{\partial f_{\text{ar}}}(f_{\text{ae}}) < 0$. As shown in~\cite{SM}, the average RP non-linearly deforms the drop as in~\cite{issenmann2006}. As a result, each time the oscillation instability is observed, the steady deformation satisfies $\frac{\partial \hat{P}}{\partial f_{\text{ar}}}(f_{\text{ae}}) > 0$. This basically explains why oscillations are also observed on the left-hand side of the peak of the $\hat{P}(f_{\text{ae}})$ vs $H$ curve. 

We note that the phase-locking mechanism proposed in~\cite{Mahravan_2016,Mahravan_2020} for explaining the instability of low-frequency surface eigenmodes  ($f_{\text{so}} \ll f_{\text{ar}}$) predicts that (i) the components along the liquid surface of the acoustic and surface eigenmodes coincide and (ii) the larger the acoustic amplitude, the larger the amplitude of the surface oscillation. This is not experimentally observed here, see Figs.~\ref{figure3},~\ref{figure4}d and~\ref{figure2}b (even the symmetries of the acoustic and surface eigenmodes even do not coincide), thus making this model inadequate for explaining our observations, possibly because it does not take into account the feedback of the drop shape on the acoustic field.

In the light of these results, first we suggest that SAW-driven surface turbulence~\cite{blamey2013} may originate from the independent excitation of numerous surface eigenmodes above different thresholds since many surface eigenmodes can in principle interact with the intracavity acoustic field, as shown in~\cite{SM}. Moreover, we note that the contact-line pinning, which usually hinders drop transport, can be overcome by low-frequency drop oscillations \cite{baudoin2012,bussonniere2016}, thus making of this oscillation instability an essential ingredient of SAW-induced transport. Indeed, since intracavity resonances are expected to persist up to $40$~MHz~\cite{brunet2010}, we expect the above unraveled instability mechanism to hold in the higher frequency range where drop transport is usually achieved, possibly cooperating with acoustic streaming. Finally, the instability can be also triggered by a hybrid combination of surface and bulk waves~\cite{rezk2016} and is therefore generic, as shown by the experiments dedicated to the imaging of the drop surface deformations presented in~\cite{SM}.

\begin{acknowledgments}
M.B. first observed the phenomenon. R.W. proposed and designed the experiments and supervised the project. N.C. built the setups and carried out the experiments and their analysis. P.B. and M.B. provided experimental support and advice. M.B. performed the numerical computations. All authors contributed to the interpretation of the results. R.W. proposed and derived the model and wrote the manuscript. M.B. and P.B. helped shape the research and manuscript.

The authors thank J.L. Thomas, A. Riaud and O. Bou-Matar for their help in the use of their vibrometer, J. Marchall for his help in instrumentation and J.L. Thomas, J. Pierre and A. Bussonni\`{e}re for helpful advices.
\end{acknowledgments}


%
\end{document}